\documentclass[11pt]{article}
\makeatletter
\@addtoreset{equation}{section}

\makeatother
\usepackage{fullpage}
\usepackage{pmat}
\usepackage{amsmath,amsthm,amsfonts,amssymb}
\usepackage{graphicx}
\usepackage[all]{xypic}

\begin{document}

\null
\begin{center}

{\Large \bf B\"acklund Transformations for Noncommutative}

 \vskip 0.2cm
 
{\Large \bf  Anti-Self-Dual Yang-Mills Equations}

\vskip 0.4cm

{\large Claire R. Gilson${}^1$, Masashi Hamanaka${}^2$ and Jonathan J. C. Nimmo${}^1$
 }

\vskip 0.2cm

        ${}^1${\it Department of Mathematics, University of Glasgow,
              Glasgow G12 8QW, UK}



        ${}^2${\it Department of Mathematics, University of Nagoya,
                      Nagoya, 464-8602, JAPAN}
%
%
%
%

\vskip 0.4cm

{\bf Abstract}
\end{center}
\baselineskip 0.5mm
{\small
We present B\"acklund transformations for
the noncommutative anti-self-dual Yang-Mills equation 
where the gauge group is $G=GL(2)$ and use it to generate a
series of exact solutions from a simple seed solution. 
The solutions generated by this approach are 
represented in terms of quasideterminants
and belong to a noncommutative version of the Atiyah-Ward ansatz.
In the commutative limit, our results coincide with
those by Corrigan, Fairlie, Yates and Goddard.
}
\baselineskip 4.3mm 

\section{Introduction}

Noncommutative (NC) extensions of integrable systems and soliton theory
have been studied intensively for the last few years
in various contexts of both mathematics and physics.
(For reviews, see e.g. \cite{integ}.)
In particular, the extension to NC spaces has drawn much attention
in physics, because in gauge theories this kind of NC extension
corresponds to the presence of background magnetic fields,
and various applications have been made successfully.
(For reviews, see e.g. \cite{NC}.)

In commutative gauge theories, the anti-self-dual Yang-Mills (ASDYM) equation
is quite important. The finite-action solutions, 
that is the {\it instanton} solutions, play key roles 
in field theories and in 4-dimensional geometry. 
In integrable systems, it is also very important
since many lower-dimensional integrable equations such as the
Korteweg-de Vries (KdV) equation and the Toda field equation 
can be derived as reductions of this ASDYM equation. 
This connection was first remarked on by Richard Ward \cite{Ward}
(known as {\it Ward's conjecture}) and is summarized elegantly
in the book of Mason and Woodhouse \cite{MaWo}
using twistor theory.

Following the work of Nekrasov and Schwarz \cite{NeSc}, the 
NC ASDYM equation has been investigated from various viewpoints.
It is actually integrable in some sense \cite{KKO}-\cite{BrMa}
and contain new solutions special to NC spaces. Furthermore,
many NC integrable equations are proved to be reductions of
NC ASDYM equation \cite{Hamanaka06_NPB}, and therefore
it is now worth studying the integrable aspects of the
NC ASDYM equation in detail and applying the results
to lower-dimensional integrable equations.

In the present paper, we give B\"acklund transformations for
NC ASDYM equation where the gauge group is $G=GL(2)$ and use it to generate a
series of exact solutions from a simple seed solution. 
The solutions generated by this approach are represented
in terms of quasideterminants \cite{GeRe}
and belong to a noncommutative version 
of the {\it Atiyah-Ward ansatz} \cite{AtWa}.
Quasideterminants appear also in the construction 
of exact soliton solutions in lower-dimensional integrable equations 
such as KdV equation \cite{EGR}-\cite{GNS} where they play the role that
determinants do in the corresponding commutative integrable systems.
In the commutative limit, our results coincide with
those by Corrigan, Fairlie, Yates and Goddard \cite{CFGY}.

We note that in our treatment, all multiplication is associative but is not assumed to be commutative. Hence the results we obtain are available not only for discussion on NC flat spaces which are realized by replacement of
all products with star-products, but also in other NC settings such as matrix or quarternion-valued ASDYM equations.

\section{Brief review of quasideterminants}

In this section, we give a brief introduction to quasideterminants,
introduced by Gelfand and Retakh \cite{GeRe},
in which a few of the key properties which play important roles
in the following sections are described. More detailed discussion is seen in the survey \cite{GGRW}.

Quasideterminants are defined in terms of inverse matrices
and we suppose the existence of all matrix inverses referred to.
Let $A=(a_{ij})$ be a $n\times n$ matrix and  $B=(b_{ij})$ be
the inverse matrix of $A$, that is, $A B=B A =1$.
Here the matrix entries belong to a noncommutative ring.
Quasideterminants of $A$ are defined formally
as the inverses of the entries in $B$:
$\vert A \vert_{ij}:=b_{ji}^{-1}$.
In the case that variables commute, this is reduced to
\begin{eqnarray}
 \vert A \vert_{ij}=
  (-1)^{i+j}\frac{\det A}{\det \tilde{A}^{ij}},
\label{limit}
\end{eqnarray}
where $\tilde{A}^{ij}$ is the matrix obtained from $A$ by
deleting the $i$-th row and the $j$-th column.

We can also write down a more explicit definition of quasideterminants.
In order to see this, let us recall the following formula
for the inverse a square $2\times 2$ block square matrix:
\begin{eqnarray*}
 \left[
 \begin{array}{cc}
  A&B \\C&d
 \end{array}
 \right]^{-1}
=\left[\begin{array}{cc}
A^{-1}+A^{-1} B S^{-1} C A^{-1}
 &-A^{-1} B S^{-1}\\
 -S^{-1} C A^{-1}
&S^{-1}
\end{array}\right],
\end{eqnarray*}
where $A$ is a square matrix, $d$ is a single element and $B$ and $C$ are column and row vectors of appropriate length and $S=d-C A^{-1} B$ is called a {\it Schur complement}. In fact this formula is valid for $A$, $B$, $C$ and $d$ in any ring not just for matrices. Thus the quasideterminant associated with the bottom right element is simply $S$. By choosing an appropriate partitioning, any entry in the inverse of a square matrix can be expressed as the inverse of a Schur complement and hence quasideterminants can also be defined recursively by:
\begin{eqnarray}
 \vert A \vert_{ij}=a_{ij}-\sum_{i^\prime (\neq i), j^\prime (\neq j)}
  a_{ii^\prime}  ((\tilde{A}^{ij})^{-1})_{i^\prime j^\prime} 
  a_{j^\prime
  j}
 =a_{ij}-\sum_{i^\prime (\neq i), j^\prime (\neq j)}
  a_{ii^\prime}  (\vert \tilde{A}^{ij}\vert_{j^\prime i^\prime })^{-1}
   a_{j^\prime j}.
\end{eqnarray}
It is sometimes convenient to use the following alternative notation in which a box is drawn about the corresponding entry in the matrix:
\begin{eqnarray}
 \vert A\vert_{ij}=
  \begin{array}{|ccccc|}
   a_{11}&\cdots &a_{1j} & \cdots& a_{1n}\\
   \vdots & & \vdots & & \vdots\\
   a_{i1}&~ & {\fbox{$a_{ij}$}}& ~& a_{in}\\
   \vdots & & \vdots & & \vdots\\
   a_{n1}& \cdots & a_{nj}&\cdots & a_{nn}
  \end{array}~.
\end{eqnarray}

Quasideterminants have various interesting properties
similar to those of determinants. Among them,
the following ones play important roles in this paper. In the block matrices given in these results, lower case letters denote single entries and upper case letters denote matrices of compatible dimensions so that the overall matrix is square. 

\begin{itemize}

 \item NC Jacobi identity \cite{GiNi}

A simple and useful special case of the NC Sylvester's Theorem
       \cite{GeRe} is
\begin{equation}\label{nc syl}
    \begin{vmatrix}
      A&B&C\\
      D&f&g\\
      E&h&\fbox{$i$}
    \end{vmatrix}=
    \begin{vmatrix}
      A&C\\
      E&\fbox{$i$}
    \end{vmatrix}-
    \begin{vmatrix}
      A&B\\
      E&\fbox{$h$}
    \end{vmatrix}
    \begin{vmatrix}
      A&B\\
      D&\fbox{$f$}
    \end{vmatrix}^{-1}
    \begin{vmatrix}
      A&C\\
      D&\fbox{$g$}
    \end{vmatrix}.
\end{equation}



\item Homological relations \cite{GeRe}

\begin{eqnarray}\label{row hom}
    \begin{vmatrix}
      A&B&C\\
      D&f&g\\
      E&\fbox{$h$}&i
    \end{vmatrix}
=  \begin{vmatrix}
      A&B&C\\
      D&f&g\\
      E&h&\fbox{$i$}
    \end{vmatrix}
    \begin{vmatrix}
      A&B&C\\
      D&f&g\\
      0&\fbox{0}&1
    \end{vmatrix},~~~
\label{col hom}
    \begin{vmatrix}
      A&B&C\\
      D&f&\fbox{$g$}\\
      E&h&i
    \end{vmatrix}
=      \begin{vmatrix}
      A&B&0\\
      D&f&\fbox{0}\\
      E&h&1
    \end{vmatrix}
    \begin{vmatrix}
      A&B&C\\
      D&f&g\\
      E&h&\fbox{$i$}
    \end{vmatrix}
\end{eqnarray}

\item A derivative formula for quasideterminants \cite{GiNi}

\begin{align}
    \begin{vmatrix}
    A&B\\
    C&\fbox{$d$}
    \end{vmatrix}'
 \label{col diff} &=
    \begin{vmatrix}
    A&B'\\
    C&\fbox{$d'$}
    \end{vmatrix}
    +\sum_{k=1}^n
    \begin{vmatrix}
    A&(A_k)'\\
    C&\fbox{$(C_k)'$}
    \end{vmatrix}
    \begin{vmatrix}
    A&B\\
    e_k^t&\fbox{$0$}
    \end{vmatrix}
,
\end{align}
where $A_k$ is the $k$th column of a matrix $A$ and  
$e_k$ is the column $n$-vector $(\delta_{ik})$ (i.e.\ 1 in the
$k$th row and 0 elsewhere). 

\end{itemize}

\section{B\"acklund transformation for the NC ASDYM equation}

In this section, we give B\"acklund transformations 
which leaves invariant the NC ASDYM equation, or equivalently, the 
NC Yang's equation for $G=GL(2)$. We also present explicit expressions, in terms of quasideterminants, for the solutions generated in this way. This transformation is a NC version of the so called {\it Corrigan-Fairlie-Yates-Goddard (CFYG) transformation} \cite{CFGY}. It generates a series of exact solutions which belong to NC version of the Atiyah-Ward ansatz \cite{AtWa} labelled by
a positive integer.

\subsection{The NC ASDYM equation and the NC Yang's equation}

Let us consider a special representation of the NC ASDYM equation for $G=GL(2)$:
\begin{eqnarray}
\label{yang}
 \partial_z(J^{-1}  \partial_{\tilde{z}} J)-\partial_w (J^{-1} \partial_{\tilde{w}} J)=0,
\end{eqnarray}
where $z,\tilde{z},w,\tilde{w}$
denote double null coordinates of $4$-dimensional space \cite{MaWo}. 
Equation (\ref{yang}) is called the {\it NC Yang's equation} 
and the gauge-invariant matrix $J$ 
is called {\it Yang's $J$-matrix}.
Gauge fields are obtained from a solution $J$ of the NC Yang's
equation via a decomposition $J=\tilde{h}^{-1} h$:
\begin{eqnarray*}
A_{z}=-(\partial_z  h) h^{-1}, ~~~  A_{w}=-(\partial_w h) h^{-1},~~~
A_{\tilde{z}}=-(\partial_{\tilde{z}}\tilde{h}) \tilde{h}^{-1}, ~~~
A_{\tilde{w}}=-(\partial_{\tilde{w}}\tilde{h}) \tilde{h}^{-1}.
\end{eqnarray*} 

In order to discuss B\"acklund transformations for
the NC Yang's equation, we parameterize
the $2\times 2$ matrix $J$ as
\begin{eqnarray}
 J=\left[\begin{array}{cc} f -g b^{-1} e&-g b^{-1}
   \\ b^{-1}  e &b^{-1}\end{array}
   \right].
\end{eqnarray}
This parameterization is always possible
when $f$ and $b$ are non-singular. In comparison with the commutative case, where only $f$ appears, in the NC setting, we need to introduce another valuable $b$.  In the commutative limit we may choose $b=f$. 

Then the NC Yang's equation (\ref{yang}) is decomposed as
\begin{eqnarray}
  \partial_{z}(f^{-1} g_{\tilde{z}}  b^{-1})-\partial_{w}(f^{-1} g_{\tilde{w}}
  b^{-1})=0,~~~
  \partial_{\tilde{z}}(b^{-1} e_z  f^{-1})
  -\partial_{\tilde{w}}(b^{-1} e_w f^{-1})&=&0,\nonumber\\
  \partial_{z}( b_{\tilde{z}} b^{-1})-
  \partial_w(b_{\tilde{w}} b^{-1})
  -e_z f^{-1} g_{\tilde{z}} b^{-1}
  +e_w f^{-1} g_{\tilde{w}} b^{-1}&=&0,\nonumber\\
  \partial_{z}(f^{-1} f_{\tilde{z}})-
  \partial_{w}(f^{-1} f_{\tilde{w}})
  -f^{-1} g_{\tilde{z}} b^{-1} e_{z}
  +f^{-1} g_{\tilde{w}} b^{-1} e_{w}&=&0,
\label{dYang}
\end{eqnarray}
where subscripts denote partial derivatives.

A gauge fixing corresponds to a 
decomposition of $J$ into matrices $h$ and $\tilde{h}$.
There is a simple and useful decomposition corresponding to
the above parameterization of $J$:
\begin{eqnarray}
 J = \tilde{h}^{-1} h=
   \left[\begin{array}{cc}1&g
   \\ 0 &b\end{array}
   \right]^{-1}
 \left[\begin{array}{cc}f&0
   \\ e&1\end{array}
   \right],~~~
 J^{-1}=
   \left[\begin{array}{cc}f&0
   \\ e &1\end{array}
   \right]^{-1}
 \left[\begin{array}{cc}1&g
   \\ 0&b\end{array}
   \right]
=
\left[\begin{array}{cc} f^{-1} &f^{-1} g \\
-e f^{-1} & b-e f^{-1} g
   \end{array}
   \right].
\end{eqnarray}
By using these formulae, we can find the gauge fields and the
field strength in terms of $b,e,f,g$.

\subsection{NC CFYG transformation}

Now we describe the NC CFYG transformation
explicitly. It is a composition of the following 
two B\"acklund transformations
for the decomposed NC Yang's equations (\ref{dYang}).

\begin{itemize}
\item $\beta$-transformation \cite{MaWo}: 
\begin{eqnarray*}
 \label{new}
   e_w^{\mbox{\scriptsize{new}}}=f^{-1} g_{\tilde{z}} b^{-1},
   e_z^{\mbox{\scriptsize{new}}}=f^{-1} g_{\tilde{w}} b^{-1},
   g_{\tilde{z}}^{\mbox{\scriptsize{new}}}=b^{-1} e_w f^{-1},
   g_{\tilde{w}}^{\mbox{\scriptsize{new}}}=b^{-1} e_z f^{-1},
   f^{\mbox{\scriptsize{new}}}=b^{-1},b^{\mbox{\scriptsize{new}}}=f^{-1}.
\end{eqnarray*}
The first four equations 
can be interpreted as integrability conditions
for the first two equations in (\ref{dYang}). We can easily check
that the last two equations in  (\ref{dYang})
are invariant under this transformation.
Also, it is clear that $\beta\circ\beta$ is the identity transformation.

\item $\gamma_0$-transformation:
\begin{eqnarray}
\left[\begin{array}{cc}
f^{\mbox{\scriptsize{new}}}&g^{\mbox{\scriptsize{new}}}\\
e^{\mbox{\scriptsize{new}}}&b^{\mbox{\scriptsize{new}}}
\end{array}\right]=
\left[\begin{array}{cc}
b&e\\g&f\end{array}\right]^{-1}
=
\left[\begin{array}{cc}
(b-ef^{-1} g)^{-1}&(g-f e^{-1} b)^{-1}\\
(e-b g^{-1} f)^{-1}&(f-g b^{-1} e)^{-1}\end{array}\right].
\label{gamma_0}
\end{eqnarray}
This follows from the fact that
the transformation $\gamma_0: J\mapsto 
J^{\mbox{\scriptsize{new}}}$ is equivalent to 
the simple conjugatation
$ J^{\mbox{\scriptsize{new}}}=C^{-1}JC,~
 C=\left[\begin{array}{cc}0&1\\1&0\end{array}\right],$
which clearly leaves the NC Yang's equation (\ref{yang}) invariant.
The relation (\ref{gamma_0}) is derived 
by comparing elements in this transformation.
It is a trivial fact that $\gamma_0\circ \gamma_0$ is the identity transformation.
\end{itemize}


\subsection{Exact NC Atiyah-Ward ansatz solutions} 


Now we construct exact solutions 
by using a chain of B\"acklund transformations from a seed solution.
Let us consider $b=e=f=g=\Delta_0^{-1}$, then
we can easily find that the decomposed NC Yang's equation
is reduced to a NC linear equation $(\partial_z\partial_{\tilde{z}}
-\partial_w\partial_{\tilde{w}})\Delta_0=0$. (We note that
for the Euclidean space, this is the NC Laplace equation because of
the reality condition $\bar{w}=-\tilde{w}$.)
Hence we can generate two series of exact solutions
$R_m$ and $R_m^\prime$
by iterating the $\beta$- and $\gamma_0$-transformations
one after the other as follows
(The seed solution $b=e=f=g=\Delta_0^{-1}$ belongs to $R_1$.):
\[
\xymatrix{
R_1\ar@{->}[r]^\alpha\ar@{<->}[d]_\beta&R_2\ar@{->}[r]^\alpha\ar@{<->}[d]_\beta&R_3\ar@{->}[r]^\alpha\ar@{<->}[d]_\beta&R_4\ar@{->}[r]\ar@{<->}[d]_\beta&\cdots\\
R'_1\ar@{->}[r]_{\alpha'}\ar@{<->}[ur]^{\gamma_0}&R'_2\ar@{->}[r]_{\alpha'}\ar@{<->}[ur]^{\gamma_0}&R'_3\ar@{->}[r]_{\alpha'}\ar@{<->}[ur]^{\gamma_0}&R'_4\ar@{->}[r]\ar@{<->}[ur]&\cdots
}
\]
where $\alpha=\gamma_0
\circ \beta:R_m \rightarrow R_{m+1}$
and $\alpha^{\prime}=\beta \circ \gamma_0: R^\prime_m \rightarrow
R^\prime_{m+1}$. These two kind of series of solutions 
in fact arise from some class of NC Atiyah-Ward ansatz.\footnote{
This is discussed in our forthcoming paper in detail \cite{GHN}.}
The explicit form of the solutions $R_m$ or $R_m^\prime$
can be represented in terms of quasideterminants
whose elements $\Delta_r$ ($r=-m+1,-m+2,\cdots,m-1$) satisfy
\begin{eqnarray}
\label{chasing}
 \frac{\partial \Delta_r}{\partial z}
= -\frac{\partial \Delta_{r+1}}{\partial \tilde{w}},~~~
 \frac{\partial \Delta_r}{\partial w}
= -\frac{\partial \Delta_{r+1}}{\partial \tilde{z}},~~~
-m+1\leq r\leq m-2~~~(m\geq 2),
\end{eqnarray}
which implies that every element $\Delta_r$ is a solution of
the NC linear equation $(\partial_z\partial_{\tilde{z}}
-\partial_w\partial_{\tilde{w}})\Delta_r=0$.
The results are as follows:

\begin{itemize}
 \item NC Atiyah-Ward ansatz solutions $R_m$

NC Atiyah-Ward ansatz solutions $R_m$
are represented by the explicit form 
of elements $b_m$, $e_m$, $f_m$, $g_m$ in $J_m$ as follows:
\begin{eqnarray*}
b_m&=&
\begin{array}{|cccc|}
\Delta_0&\Delta_{-1} & \cdots & \Delta_{1-m}\\
\Delta_1 &\Delta_0&\cdots & \Delta_{2-m} \\
\vdots &\vdots &\ddots & \vdots\\
\Delta_{m-1} &\Delta_{m-2} &\cdots &\fbox{$\Delta_0$} 
\end{array}^{-1},~~~
f_m=
\begin{array}{|cccc|}
\fbox{$\Delta_0$}&\Delta_{-1} & \cdots & \Delta_{1-m}\\
\Delta_1 &\Delta_0&\cdots & \Delta_{2-m} \\
\vdots &\vdots &\ddots & \vdots\\
\Delta_{m-1} &\Delta_{m-2} &\cdots &\Delta_0
\end{array}^{-1},\nonumber\\
e_m&=&
\begin{array}{|cccc|}
\Delta_0&\Delta_{-1} & \cdots & \fbox{$\Delta_{1-m}$}\\
\Delta_1 &\Delta_0&\cdots & \Delta_{2-m} \\
\vdots &\vdots &\ddots & \vdots\\
\Delta_{m-1} &\Delta_{m-2} &\cdots &\Delta_0
\end{array}^{-1},~~~
g_m
=\begin{array}{|cccc|}
\Delta_0&\Delta_{-1} & \cdots & \Delta_{1-m}\\
\Delta_1 &\Delta_0&\cdots & \Delta_{2-m} \\
\vdots &\vdots &\ddots & \vdots\\
\fbox{$\Delta_{m-1}$} &\Delta_{m-2} &\cdots &\Delta_0 
\end{array}^{-1}.
\end{eqnarray*}
In the commutative limit, we can easily see that $b_m=f_m$.
The ansatz $R_1$ 
leads to so called the {\it Corrigan-Fairlie-'t Hooft-Wilczek} 
(CFtHW) ansatz \cite{CFtHW}.

\item NC Atiyah-Ward ansatz solutions $R^\prime_m$

NC Atiyah-Ward ansatz solutions $R'_m$
are represented by the explicit form 
of elements $b'_m$, $e'_m$, $f'_m$, $g'_m$
in $\tilde{J}_m$ as follows:
\begin{eqnarray*}
b^\prime_m
&=&
\begin{array}{|cccc|}
\fbox{$\Delta_0$}&\Delta_{-1} & \cdots & \Delta_{1-m}\\
\Delta_1 &\Delta_0&\cdots & \Delta_{2-m} \\
\vdots &\vdots &\ddots & \vdots\\
\Delta_{m-1} &\Delta_{m-2} &\cdots &\Delta_0 
\end{array}~,~~~
f^\prime_m
=
\begin{array}{|cccc|}
\Delta_0&\Delta_{-1} & \cdots & \Delta_{1-m}\\
\Delta_1 &\Delta_0&\cdots & \Delta_{2-m} \\
\vdots &\vdots &\ddots & \vdots\\
\Delta_{m-1} &\Delta_{m-2} &\cdots &\fbox{$\Delta_0$}
\end{array}~,\nonumber\\
e^\prime_m
&=&
\begin{array}{|cccc|}
\Delta_{-1}&\Delta_{-2} & \cdots & \fbox{$\Delta_{-m}$}\\
\Delta_0 &\Delta_{-1}&\cdots & \Delta_{1-m} \\
\vdots &\vdots &\ddots & \vdots\\
\Delta_{m-2} &\Delta_{m-3} &\cdots &\Delta_{-1}
\end{array}~,~~~
g^\prime_m
=
\begin{array}{|cccc|}
\Delta_1&\Delta_{0} & \cdots & \Delta_{2-m}\\
\Delta_2 &\Delta_1&\cdots & \Delta_{3-m} \\
\vdots &\vdots &\ddots & \vdots\\
\fbox{$\Delta_{m}$} &\Delta_{m-1} &\cdots &\Delta_1 
\end{array}~.
\end{eqnarray*}
In the commutative case, 
$b^\prime_m=f^\prime_m$ also holds.
For $m=1$, we get $b^\prime_1=f^\prime_1=\Delta_0, e^\prime_1=\Delta_{-1}, 
g^\prime_1=\Delta_1$ 
and then the relation (\ref{chasing}) implies that 
$e^\prime_{1,z}=f^\prime_{1,\tilde{w}},~
e^\prime_{1,w}=f^\prime_{1,\tilde{z}},~
f^\prime_{1,z}= g^\prime_{1, \tilde{w}},~
f^\prime_{1, w}= g^\prime_{1, \tilde{z}},$
and leads to the CFtHW ansatz
which was first pointed out by Yang \cite{Yang}.


We can also present a compact form of the whole Yang's matrix $J$
in terms of a single quasideterminants expanded by a $2\times2$ submatrix:
\begin{eqnarray}
\label{J_q}
J_m^{\prime}=
\left[
\begin{array}{cc}
f'_m-g'_mb_m^{\prime-1}e'_m&-g'_mb_m^{\prime-1}\\
b_m^{\prime-1}e'_m&b_m^{\prime-1}
\end{array}\right]=
\begin{pmat}|{|..||}|
\Delta_0&\Delta_{-1}&\cdots&\Delta_{1-m}&\Delta_{-m}&-1\cr\-
\Delta_1&\Delta_0&\cdots&\Delta_{2-m}&\Delta_{1-m}&0\cr
\vdots&\vdots&\ddots&\vdots&\vdots&\vdots\cr
\Delta_{m-1}&\Delta_{m-2}&\cdots&\Delta_0&\Delta_{-1}&0\cr\-
\Delta_{m}&\Delta_{m-1}&\cdots&\Delta_1&\fbox{$\Delta_{0}$}&\fbox{0}\cr\-
1&0&\cdots&0&\fbox{0}&\fbox{0}\cr
\end{pmat}.
\end{eqnarray}
and
\begin{eqnarray}
\label{J^-1_q}
J_m^{\prime -1}=\left[
\begin{array}{cc}
f_m^{\prime-1}&f_m^{\prime-1}g'_m\\
-e'_mf_m^{\prime-1}&b'_m-e'_mf_m^{\prime-1}g'_m
\end{array}\right]
=
-\begin{pmat}|{||..|}|
\fbox{0}&\fbox{0}&0&\cdots&0&1\cr\-
\fbox{0}&\fbox{$\Delta_0$}&\Delta_{-1}&\cdots&\Delta_{1-m}&\Delta_{-m}\cr\-
0&\Delta_1&\Delta_0&\cdots&\Delta_{2-m}&\Delta_{1-m}\cr
\vdots&\vdots&\vdots&\ddots&\vdots&\vdots\cr
0&\Delta_{m-1}&\Delta_{m-2}&\cdots&\Delta_0&\Delta_{-1}\cr\-
-1&\Delta_{m}&\Delta_{m-1}&\cdots&\Delta_1&\Delta_{0}\cr
\end{pmat}.
\end{eqnarray}
Because $J$ is gauge invariant, this shows that
the present B\"acklund transformation is not
just a gauge transformation but a non-trivial one.
The proof of these representations is given in Appendix A
and in \cite{GiGu}.
\end{itemize}

The $\gamma_0$-transformation
is proved simply using NC Jacobi identity (\ref{nc syl}) 
applied to the four corner elements.\footnote{We note that
quasideterminants are invariant under any permutations of the rows and columns 
\cite{GeRe}.}
For example,
\begin{eqnarray}
b_{m+1}^{-1}
&=&
\begin{array}{|ccc|}
\Delta_0&\cdots & \Delta_{1-m}\\
\vdots &\ddots & \vdots\\
\Delta_{m-1} &\cdots &\fbox{$\Delta_0$} 
\end{array}
-\begin{array}{|ccc|}
\Delta_1&\cdots & \Delta_{2-m}\\
\vdots &\ddots & \vdots\\
\fbox{$\Delta_{m}$} &\cdots &\Delta_1 
\end{array}~
\begin{array}{|ccc|}
\fbox{$\Delta_0$}&\cdots & \Delta_{1-m}\\
\vdots &\ddots & \vdots\\
\Delta_{m-1} &\cdots &\Delta_0 
\end{array}^{-1}
\begin{array}{|ccc|}
\Delta_{-1} &\cdots & \fbox{$\Delta_{-m}$} \\
\vdots &\ddots & \vdots\\
\Delta_{m-2} &\cdots &\Delta_{-1} 
\end{array}
\nonumber\\  
&=&f^\prime_m-g^\prime_m b^{\prime -1}_m
e^\prime_m.
\end{eqnarray}

The proof of the $\beta$-transformation
uses both the NC Jacobi identity and also 
the homological relations \eqref{row hom}. 
We will prove the first equation in the $\beta$-transformation,
that is, if the $\Delta_i$
satisfy $\Delta_{i,w}=-\Delta_{i+1,\tilde z}$ then
\[
	e'_{m,w}=f_m^{-1}g_{m,\tilde z}b_m^{-1}.
\]

The RHS is equal to
\[
	-b_m'g_m(g_m^{-1})_{\tilde z}g_mf_m'.
\]
In this, it follows from \eqref{row hom} that the first two and last two factors are
\begin{eqnarray}
b_m'g_m=
\begin{vmatrix}
\fbox{$0$}&\Delta_{-1}&\cdots&\Delta_{1-m}\\
0&\Delta_{0}&\cdots&\Delta_{2-m}\\
\vdots&\vdots&&\vdots\\
0&\Delta_{m-3}&\cdots&\Delta_{-1}\\
1&\Delta_{m-2}&\cdots&\Delta_{0}
\end{vmatrix},~~~
g_mf_m'=
\begin{vmatrix}
\Delta_{0}&\Delta_{-1}&\cdots &\Delta_{2-m} &\Delta_{1-m}\\
\vdots&\vdots&&\vdots&\vdots\\
\Delta_{m-2}&\Delta_{m-3}&\cdots&\Delta_{0}&\Delta_{-1}\\
1&0&\cdots&0&\fbox{$0$}
\end{vmatrix}.
\end{eqnarray}

Next, from \eqref{col diff}, we have 
\[
\begin{split}
&(g_m^{-1})_{\tilde z}=
\begin{vmatrix}
\Delta_{0,\tilde z}&\Delta_{-1}&\cdots&\Delta_{1-m}\\
\Delta_{1,\tilde z}&\Delta_{0}&\cdots&\Delta_{2-m}\\
\vdots&\vdots&&\vdots\\
\Delta_{m-2,\tilde z}&\Delta_{m-3}&\cdots&\Delta_{-1}\\
\fbox{$\Delta_{m-1,\tilde z}$}&\Delta_{m-2}&\cdots&\Delta_{0}
\end{vmatrix}\\&+\sum_{k=1}^{m-1}
\begin{vmatrix}
\Delta_{-k,\tilde z}&\Delta_{-1}&\cdots&\Delta_{1-m}\\
\Delta_{1-k,\tilde z}&\Delta_{0}&\cdots&\Delta_{2-m}\\
\vdots&\vdots&&\vdots\\
\Delta_{m-2-k,\tilde z}&\Delta_{m-3}&\cdots&\Delta_{-1}\\
\fbox{$\Delta_{m-1-k,\tilde z}$}&\Delta_{m-2}&\cdots&\Delta_{0}
\end{vmatrix}
\begin{vmatrix}
\Delta_{0}&\Delta_{-1}&\cdots &\Delta_{-k} &\cdots&\Delta_{1-m}\\
\vdots&\vdots&&\vdots&&\vdots\\
\Delta_{m-2}&\Delta_{m-3}&\cdots&\Delta_{m-2-k}&\cdots&\Delta_{-1}\\
\fbox{$0$}&0&\cdots&1&\cdots&0
\end{vmatrix}.
\end{split}
\]
The effect of the left and right factors
on this expression is to move expansion points
as specified in \eqref{row hom} 
and so, after moving the expansion column (row)
in the first (second) factors to the rightmost (top), we get
\[
\begin{split}
&f_m^{-1}g_{m,\tilde z}b_m^{-1}=
-\begin{vmatrix}
\Delta_{-1}&\cdots&\Delta_{1-m}&\fbox{$\Delta_{1-m,\tilde z}$}\\
\Delta_{0}&\cdots&\Delta_{2-m}&\Delta_{2-m,\tilde z}\\
\vdots&&\vdots&\vdots\\
\Delta_{m-3}&\cdots&\Delta_{-1}&\Delta_{-1,\tilde z}\\
\Delta_{m-2}&\cdots&\Delta_{0}&\Delta_{0,\tilde z}
\end{vmatrix}\\&
-\sum_{k=0}^{m-2}
\begin{vmatrix}
\Delta_{-1}&\cdots&\Delta_{1-m}&\fbox{$\Delta_{-k,\tilde z}$}\\
\Delta_{0}&\cdots&\Delta_{2-m}&\Delta_{1-k,\tilde z}\\
\vdots&&\vdots&\vdots\\
\Delta_{m-2}&\cdots&\Delta_{0}&\Delta_{m-1-k,\tilde z}
\end{vmatrix}
\begin{vmatrix}
0&\cdots&1&\cdots&0&\fbox{$0$}\\
\Delta_{0}&\cdots &\Delta_{-k} &\cdots&\Delta_{2-m}&\Delta_{1-m}\\
\vdots&&\vdots&&\vdots&\vdots\\
\Delta_{m-2}&\cdots&\Delta_{m-2-k}&\cdots&\Delta_{0}&\Delta_{-1}
\end{vmatrix}.
\end{split}
\]
On the other hand, 
\[
\begin{split}
&e'_{m,w}=
\begin{vmatrix}
\Delta_{-1}&\cdots&\Delta_{1-m}&\fbox{$\Delta_{-m,w}$}\\
\Delta_{0}&\cdots&\Delta_{2-m}&\Delta_{1-m,w}\\
\vdots&&\vdots&\vdots\\
\Delta_{m-3}&\cdots&\Delta_{-1}&\Delta_{-2,w}\\
\Delta_{m-2}&\cdots&\Delta_{0}&\Delta_{-1,w}
\end{vmatrix}\\&+\sum_{k=0}^{m-2}
\begin{vmatrix}
\Delta_{-1}&\cdots&\Delta_{1-m}&\fbox{$\Delta_{-k-1,w}$}\\
\Delta_{0}&\cdots&\Delta_{2-m}&\Delta_{-k,w}\\
\vdots&&\vdots&\vdots\\
\Delta_{m-2}&\cdots&\Delta_{0}&\Delta_{m-2-k,w}
\end{vmatrix}
\begin{vmatrix}
0&\cdots&1&\cdots&0&\fbox{$0$}\\
\Delta_{0}&\cdots &\Delta_{-k} &\cdots&\Delta_{2-m}&\Delta_{1-m}\\
\vdots&&\vdots&&\vdots&\vdots\\
\Delta_{m-2}&\cdots&\Delta_{m-2-k}&\cdots&\Delta_{0}&\Delta_{-1}
\end{vmatrix},
\end{split}
\]
and then the result follows immediately from \eqref{chasing}.

\section{Conclusion and Discussion}
In this paper, we have presented B\"acklund transformations for the
NC ASDYM equation with $G=GL(2)$ and constructed from a simple seed solution a
series of exact NC Atiyah-Ward ansatz solutions expressed explicitly
in terms of quasideterminants.

In this short paper we have not discussed several important points
concerning the NC CFYG transformation, for example, the origin of
this transformation in the framework of NC twistor theory,
analysis of explicit exact solutions, the relationship with NC Darboux
and NC binary Darboux transformation \cite{HSS07} and so on.
These are reported in our forthcoming more detailed paper \cite{GHN}.
NC extension of a bilinear form approach to the ASDYM equation \cite{MOS}
is also interesting topics because many aspects in their paper
are close to ours.

These results could be applied also to lower-dimensional systems 
via the results on the NC Ward's conjecture
including NC monopoles, NC KdV equations and so on,
and might shed light
on a profound connection between higher-dimensional integrable
systems related to twistor theory and lower-dimensional ones
related to Sato's theory.

\subsection*{Acknowledgments}

MH would like to thank to L.~Mason 
for a lot of helpful comments
and hospitality during stay at Mathematical Institute,
University of Oxford, 
and to JJCN for hospitality during stay at Department of Mathematics,
University of Glasgow,
and to organizers for hospitality during the ISLAND3 conference
in Islay. The work of MH was supported by
the Yamada Science Foundation
for the promotion of the natural science
and Grant-in-Aid for Young Scientists (\#18740142).

\begin{appendix}
\section{Proof of the compact form for the NC 
Atiyah-Ward ansatz solutions}

Consider the $2\times 2$ matrices of quasideterminants of the form
\begin{align*}
\begin{vmatrix}
a&B&c&\alpha\\
D&E&F&0\\
g&H&\fbox{\hphantom0\llap{$i$}}&\fbox{0}\\
\beta&0&\fbox{0}&\fbox{0}\\
\end{vmatrix}&
:=
\begin{bmatrix}
\begin{vmatrix}
a&B&c\\
D&E&F\\
g&H&\fbox{\hphantom0\llap{$i$}}
\end{vmatrix}&\begin{vmatrix}
a&B&\alpha\\
D&E&0\\
g&H&\fbox{0}
\end{vmatrix}
\\
\begin{vmatrix}
a&B&c\\
D&E&F\\
\beta&0&\fbox{0}\\
\end{vmatrix}&
\begin{vmatrix}
a&B&\alpha\\
D&E&0\\
\beta&0&\fbox{0}\\
\end{vmatrix}
\end{bmatrix}
=
\begin{bmatrix}
\begin{vmatrix}
a&B&c\\
D&E&F\\
g&H&\fbox{\hphantom0\llap{$i$}}
\end{vmatrix}&\alpha\begin{vmatrix}
a&B&1\\
D&E&0\\
g&H&\fbox{0}
\end{vmatrix}
\\
\beta\begin{vmatrix}
a&B&c\\
D&E&F\\
1&0&\fbox{0}\\
\end{vmatrix}&
\alpha\beta\begin{vmatrix}
a&B&1\\
D&E&0\\1&0&\fbox{0}\\
\end{vmatrix}
\end{bmatrix}
,\\
\begin{vmatrix}
\fbox{0}&\fbox{0}&0&\gamma\\
\fbox{0}&\fbox{\hphantom0\llap{$a$}}&B&c\\
0&D&E&F\\
\delta&g&H&i
\end{vmatrix}&
:=
\begin{bmatrix}
\begin{vmatrix}
\fbox{0}&0&\gamma\\
0&E&F\\
\delta&H&i
\end{vmatrix}&
\begin{vmatrix}
\fbox{0}&0&\gamma\\
D&E&f\\
g&H&i
\end{vmatrix}\\
\begin{vmatrix}
\fbox{0}&B&c\\
0&E&F\\
\delta&H&i
\end{vmatrix}&
\begin{vmatrix}
\fbox{\hphantom0\llap{$a$}}&B&c\\
D&E&F\\
g&H&i
\end{vmatrix}
\end{bmatrix}
=
\begin{bmatrix}
\gamma\delta
\begin{vmatrix}
\fbox{0}&0&1\\
0&E&F\\
1&H&i
\end{vmatrix}&
\gamma
\begin{vmatrix}
\fbox{0}&0&1\\
D&E&f\\
g&H&i
\end{vmatrix}\\
\delta
\begin{vmatrix}
\fbox{0}&B&c\\
0&E&F\\
1&H&i
\end{vmatrix}&
\begin{vmatrix}
\fbox{\hphantom0\llap{$a$}}&B&c\\
D&E&F\\
g&H&i
\end{vmatrix}
\end{bmatrix},
\end{align*}
where lower case letters denote single entries, upper case letters
denote matrices of compatible dimensions and Greek letters are
scalars (i.e.{} commute with everything). Using \eqref{nc syl}
these can be rewritten as
\begin{equation*}
\begin{bmatrix}
\begin{vmatrix}
E&F\\
H&\fbox{$i$}
\end{vmatrix}-\begin{vmatrix}
D&E\\
\fbox{$g$}&H
\end{vmatrix}\begin{vmatrix}
\fbox{$a$}&B\\
D&E
\end{vmatrix}^{-1}
\begin{vmatrix}
B&\fbox{$c$}\\
E&F
\end{vmatrix}
& -\alpha
\begin{vmatrix}
D&E\\
\fbox{$g$}&H
\end{vmatrix}\begin{vmatrix}
\fbox{$a$}&B\\
D&E
\end{vmatrix}^{-1}\\-\beta
\begin{vmatrix}
\fbox{$a$}&B\\
D&E
\end{vmatrix}^{-1}\begin{vmatrix}
B&\fbox{$c$}\\
E&F
\end{vmatrix}&
-\alpha\beta
\begin{vmatrix}
\fbox{$a$}&B\\
D&E\\
\end{vmatrix}^{-1}
\end{bmatrix}
\end{equation*}
and
\begin{equation*}
\begin{bmatrix}
-\gamma\delta\begin{vmatrix}
E&F\\
H&\fbox{$i$}
\end{vmatrix}^{-1}
&-\gamma\begin{vmatrix}
E&F\\
H&\fbox{$i$}
\end{vmatrix}^{-1}
\begin{vmatrix}
D&E\\
\fbox{$g$}&H
\end{vmatrix}\\-\delta
\begin{vmatrix}
B&\fbox{$c$}\\
E&F
\end{vmatrix}\begin{vmatrix}
E&F\\
H&\fbox{$i$}
\end{vmatrix}^{-1}&\begin{vmatrix}
\fbox{$a$}&B\\
D&E
\end{vmatrix}-
\begin{vmatrix}
B&\fbox{$c$}\\
E&F
\end{vmatrix}\begin{vmatrix}
E&F\\
H&\fbox{$i$}
\end{vmatrix}^{-1}\begin{vmatrix}
D&E\\
\fbox{$g$}&H
\end{vmatrix}
\end{bmatrix}.
\end{equation*}
The product of these matrices is $I$ if and only if
$\alpha\beta=\gamma\delta=-1$ and $\alpha+\gamma=0$ and so we choose
$\alpha=-\beta=-\gamma=\delta=1$.
Thus
\begin{equation}\label{q id}
\begin{vmatrix}
a&B&c&1\\
D&E&F&0\\
g&H&\fbox{\hphantom0\llap{$i$}}&\fbox{0}\\
-1&0&\fbox{0}&\fbox{0}\\
\end{vmatrix}=\begin{bmatrix}
\begin{vmatrix}
E&F\\
H&\fbox{$i$}
\end{vmatrix}-\begin{vmatrix}
D&E\\
\fbox{$g$}&H
\end{vmatrix}\begin{vmatrix}
\fbox{$a$}&B\\
D&E
\end{vmatrix}^{-1}
\begin{vmatrix}
B&\fbox{$c$}\\
E&F
\end{vmatrix}
&
-\begin{vmatrix}
D&E\\
\fbox{$g$}&H
\end{vmatrix}\begin{vmatrix}
\fbox{$a$}&B\\
D&E
\end{vmatrix}^{-1}\\
\begin{vmatrix}
\fbox{$a$}&B\\
D&E
\end{vmatrix}^{-1}\begin{vmatrix}
B&\fbox{$c$}\\
E&F
\end{vmatrix}&
\begin{vmatrix}
\fbox{$a$}&B\\
D&E\\
\end{vmatrix}^{-1}
\end{bmatrix}
\end{equation}
and
\begin{equation}\label{q id2}\begin{vmatrix}
\fbox{0}&\fbox{0}&0&-1\\
\fbox{0}&\fbox{\hphantom0\llap{$a$}}&B&c\\
0&D&E&F\\
1&g&H&i
\end{vmatrix}=
\begin{bmatrix}
\begin{vmatrix}
E&F\\
H&\fbox{$i$}
\end{vmatrix}^{-1}
&\begin{vmatrix}
E&F\\
H&\fbox{$i$}
\end{vmatrix}^{-1}
\begin{vmatrix}
D&E\\
\fbox{$g$}&H
\end{vmatrix}\\
-\begin{vmatrix}
B&\fbox{$c$}\\
E&F
\end{vmatrix}\begin{vmatrix}
E&F\\
H&\fbox{$i$}
\end{vmatrix}^{-1}&\begin{vmatrix}
\fbox{$a$}&B\\
D&E
\end{vmatrix}-
\begin{vmatrix}
B&\fbox{$c$}\\
E&F
\end{vmatrix}\begin{vmatrix}
E&F\\
H&\fbox{$i$}
\end{vmatrix}^{-1}\begin{vmatrix}
D&E\\
\fbox{$g$}&H
\end{vmatrix}
\end{bmatrix},
\end{equation}
are inverse to each other.

To apply these results to the Atiyah-Ward ansatz solutions of $R_m^\prime$,
we get the representation (\ref{J_q}) and (\ref{J^-1_q}).

\end{appendix}

\end{document}